# TUnit – Unit Testing For Template-based Code Generators

Carsten Kolassa[1]   Markus Look[1], Klaus Müller[1], Alexander Roth[1], Dirk Reiß[2], Bernhard Rumpe[1]

**Abstract:** Template-based code generator development as part of model-driven development (MDD) demands for strong mechanisms and tools that support developers to improve robustness, i.e., the desired code is generated for the specified inputs. Although different testing methods have been proposed, a method for testing only parts of template-based code generators that can be employed in the early stage of development is lacking. Thus, in this paper we present an approach and an implementation based on JUnit to test template-based code generators. Rather than testing a complete code generator, it facilitates partial testing by supporting the execution of templates with a mocked environment. This eases testing of code generators in early stages of development as well as testing new or changed parts of a code generator. To test the source code generated by the templates under test, different methods are presented including string comparisons, API-based assertions, and abstract syntax tree based assertions.

## 1   Introduction

With the increasing adoption of model-driven development (MDD) in research and industry [Hu11, Li14], code generation - systematic transformation of compact models to detailed code [FR07] - is gaining importance. To support code generator developers in constructing robust code generators, i.e., code generators that produce the desired code for the specified input, sophisticated mechanisms and tools are required. They have to be integrable into the development process and especially for agile development processes they need to enable partial testing of code generators. Such testing as an essential activity is, however, challenging [St07].

Current approaches for testing code generators (cf. [St07, Jö13, SWC05, Ra10, St06]) require an initial integration effort of the testing procedure, are based on string comparisons only, or are designed to test the code generator as a whole. Other code generator testing approaches employ formal methods [BKS04]. Setting up such tests is time-consuming and once set up they are hard to maintain in an evolving environment, because small changes in the code generator may lead to larger changes in the tests. Consequently, existing testing approaches for code generators are not so easy to use in an agile development environment, where either the code generator does not yet generate complete code artifacts that the tests can validate or only the increment in functionality is to be tested. In summary, exisiting work lacks approaches for partial code generator testing.

---
[1] RWTH Aachen University, Software Engineering, Germany, http://www.se-rwth.de
[2] TU Braunschweig, Institute for Building Services and Energy Design, Germany

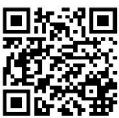

Focusing on template-based code generation, the goal of this paper is to present an approach for testing individual components of template-based code generators that can be employed in early development stage of code generators. We introduce TUnit, an extension of JUnit [JU15] based on the MontiCore [KRV10, Gr08, KRV08] language workbench to support unit testing of code generator templates. In our case, testing a code generator or parts of it means to answer the following questions: *Is the set of specified inputs accepted by the code generator template, e.g., code is generated? Does the code generator template produce syntactically valid source code? Are the target language context conditions valid for the generated source code?* Executing a TUnit test case will run the template under test with a mocked context (e.g. mocked variables, mocked templates, or mocked helper functionality) on (parts of) an input model. This approach allows for testing the output of a single template under test that is part of the overall output of the code generator, rather than testing the whole output of a code generator run. To validate that the template output meets the testers expectations, TUnit provides different kinds of assertion mechanisms including abstract syntax based comparisons and abstract syntax API-based assertions. Additionally, because string comparisons are widely used and sometimes practical, TUnit provides support for such comparisons as well. However, this approach is not robust, as the template output can change on a regular basis, e.g., due to new or deleted whitespaces.

The contributions of this paper are: (a) an understanding of a template engine context (b) concepts for mocking a template's context with nested templates to allow for partial code generator testing in early stage of the development cycle, (c) concepts for abstract syntax based testing of the partial generated source code, and (d) an implementation of these concepts within a widely used testing framework.

The paper is structured as follows: at first, we present an overview of related work (Section 2) and point out their shortcomings. Next, we introduce MontiCore (Section 3), a framework for language processing and code generation, that has been used to implement parts of TUnit. By starting with a basic TUnit test, we point out how template-based code generators can be unit-tested and which challenges need to be solved (Section 4). These challenges are addressed in Section 5. Finally, we conclude our paper in Section 6.

## 2  Related Work

With the emerging importance of MDD, code generation has received growing attention. In order to support code generator development and ensure code generator robustness, different code generator testing approaches have been proposed and are presented in more detail in [SWC05]. In the remainder of this section, we point out the main ideas of the different testing approaches that target testing of complete code generators.

CoGenTe is a tool for testing code generators [Ra10]. It takes a syntactic and a semantic meta-model of the input language and a test specification, which is a coverage criterion over the meta-model. A generator creates a test-suite that can test any code generator for the particular input language. The generated test-suite is derived using a constraint generator, an inference tree generator, and a constraint solver. Each test-suite comprises several input models and expected outputs in the target language. To test a code generator, the test-suite input models are passed to the code generator and the generated output is compared to the expected output of the test-suite. In contrast to this approach, we present an approach to test parts of a code generator for predefined input models.

Another approach to test code generators has been proposed in [St06, St07]. It is based on a formal specification of the code generator transformation as a graph rewriting rule and comprises three steps. In the first step - model-in-the-loop - the test model is transformed into an executable model that is simulated. In the second step - software-in-the-loop - the generated model is transformed by the code generator into executable code. Both, the execution results of the simulated model and the execution results of the executed code are finally compared. Existing approaches can be applied to extend this approach by automatically generating the input test-cases [Ze06, Sa08]. In contrast to this technique, our approach uses an instance of the input langue rather than a formal specification. Furthermore, no intermediate model is used for simulation. Our proposed approach works directly on the input model and not only strings but also abstract syntax trees (ASTs) can be compared. Furthermore, an AST-based API is provided that allows to check the generated output.

An instance of the above code generator testing approach to generate JUnit tests has been proposed in [Jö13]. Code generators are modeled as services from atomic service independent building blocks (SIBs). Such SIBs are used to model test cases, which are part of test suites. A code generator transforms the test cases into JUnit test scripts. The execution footprint - basically a string of the SIBs that have been executed - of direct execution of the test data and the execution footprint of the generated and compiled code are compared. The test is successful if the footprints are equal. In this paper, we focus on partial testing of code generators and use, e.g., AST comparisons for validating the generated output.

## 3    Language Processing and Code Generation with MontiCore

The MontiCore framework [KRV10, Gr08, KRV08] is the foundation for all aspects of language definition, language processing, and template-based code generation in TUnit. In the remainder of this paper, we regard a model as an instance of a language that is processed by the MontiCore framework and used for code generation. The basic structure of the MontiCore framework is shown in Figure 1. The components depicted in the upper left corner including `Grammar`, `MontiCore`, `Symboltable Entries`, `Model`, `Parser+Infrastructure` and `AST` are used for language definition and language processing, i.e., processing an input model. All other components and the components `Symboltable Entries` and `AST` are used for code generation.

The MontiCore framework uses a grammar defining the language to be processed and generates a parser and infrastructure for language processing, which are used to parse models. Each input model needs to conform to the grammar. When reading and processing models, the parser creates an AST that represents their internal structure. This abstract representation of the input model is used for both: further language processing steps and code generation. Besides the AST, MontiCore uses the `Symboltable Entries` component to create symbol table entries for each symbol of the processed models. Each symbol table entry contains information about the model structure, an element's name, and context information. This stored information is used for referencing symbols in different models and can be used to extend the language processing by defining constraints for the input model or for code generation to retrieve additional information on model symbols.

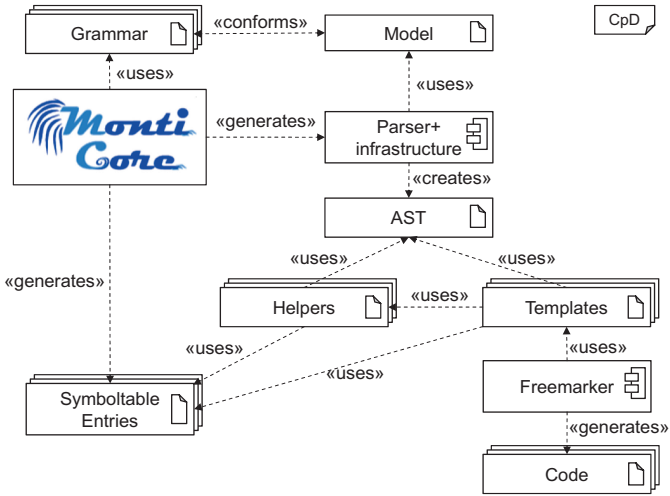

Fig. 1: Overview of MontiCore components for language processing and code generation [Sc12].

### 3.1 Template-based Code Generation with MontiCore

The MontiCore code generation process is based on a template mechanism. Templates written in FreeMarker [Fr15] describe what is to be generated. These templates, which are hierarchically structured via sub-templates, contain target code and FreeMarker expressions that finally produce target code. An overview of a template and its context is depicted in Figure 2. The result of the code generation process is the actual output labeled generated code in the figure.

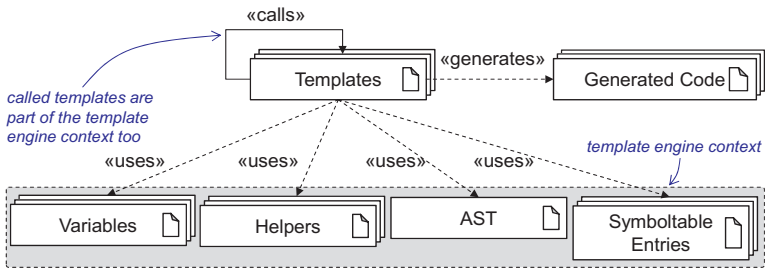

Fig. 2: Elements of the template engine context for template-based code generation in MontiCore.

Each code generation process is started by invoking a root template, which usually calls one or more sub-templates. A template has access to its template engine context object, which contains variables, helper objects, called templates, and symbol table entries, as depicted in Figure 2. A simplified example of a FreeMarker template is given in List. 1. For a better presentation of the template engine context, this listing shows two variables, the ast, and one helper.

```
                                                                    FreeMarker
1 //Variables: paramType, paramName
2 //Helpers: methodHelper
3 public ${ast.returnType} ${ast.name}
4    (${paramType} ${paramName})
5    ${methodHelper.printThrowsDecl(ast)}
```

List. 1: The (simplified) template for generating a Java method.

The primary input for templates is the AST which is constructed by processing a model file. For presentational reasons, we primarily focus on class diagrams as an input model, i.e., the AST describes the abstract syntax of a class diagram and AST elements are elements from class diagrams including classes, associations, methods, interfaces, and enumerations [Sc12]. A template is called with an AST element of the corresponding model and, in our case, generates Java source code. This AST element can be accessed through the context variable `ast` as shown in an excerpt of a template in List. 1. In this listing, the method `name` of the corresponding AST class is invoked in line 3 to return the name of the model element which is represented by the AST element. This is denoted by the FreeMarker specific syntax ${...}. Additionally, the template excerpt in List. 1 shows how variables (`paramName` and `paramType`) and helpers (`methodHelper`) are used. The meaning of variables and helpers is explained in more detail in the following.

A template may define local variables, which can be used and modified inside the template. The value for each variable is set when the template is called. For example, the template outlined in List. 1 expects that the values for the variables `paramName` and `paramType` are set when the template is called. The values of these variables are accessed in line 4 to introduce the name and the type of the method parameter into the generated code.

According to the principle of separation of concerns, templates contain target code and simple computations including string concatenations, loops and if-else conditions. In addition to that, further functionality can be implemented in helper classes in Java which are invoked from templates. When a template is called, an instance of the helper class is passed to the template and can be accessed through a helper variable. In List. 1, `methodHelper` in line 5 is a helper variable and it is used to invoke the helper method `printThrowsDecl` which returns the Java throws declaration of the method. The AST is a special kind of helper variable, as it can be used to invoke specific methods from the AST classes.

In order to test templates in isolation, we need to be able to replace either all or only some of the variables and helpers in a template's context with mocked ones. For instance, it might be desirable to apply a specific mock helper class instead of the helper class which would be used by default. Or it might be desired to set the variables to specific values.

## 4 Code Generator Template Testing with TUnit

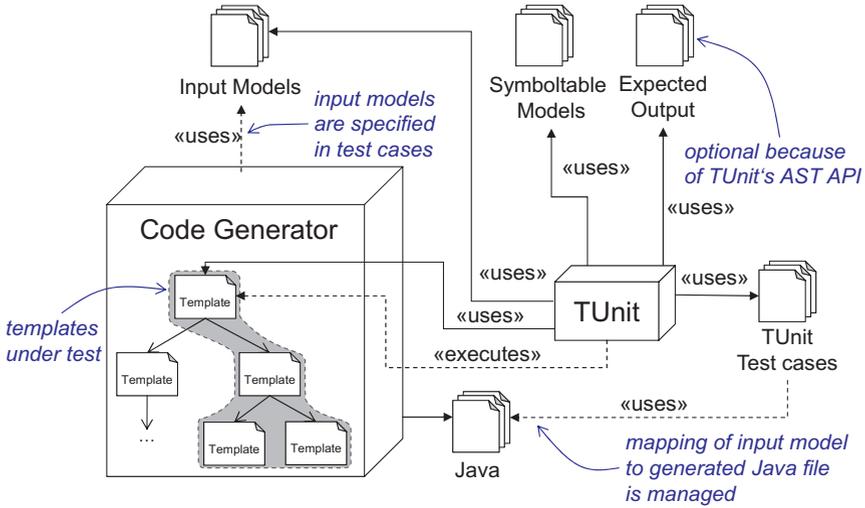

Fig. 3: Overview of our code generator testing approach.

A template-based code generator typically comprises multiple code generator templates in order to generate source code from an input model. For ease of presentation we assume that a code generator accepts one class diagram as input and generates Java source code. In addition, each code generator has access to a symbol table, where the symbols of all symboltable models are stored to identify referenced symbols. Figure 3 depicts an overview of our white-box approach to test code generator templates. In our approach TUnit test cases define the tests, symboltable models used in tests, and the templates under test. Using this input, the templates under test are executed but only for the elements of the model that have been predefined in the TUnit test case. The generated output may then either be compared to the expected output or the TUnit's AST API can be used to define assertions.

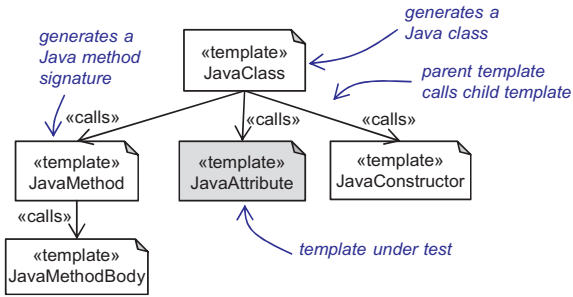

Fig. 4: An example of a code generator's template hierarchy to generate a Java class.

To give a more detailed understanding of how our testing approach works, we assume that the code generator's templates are structured as depicted in Figure 4 and that we want to test the `JavaAttribute` template from the template hierarchy. Its FreeMarker source code is listed in List. 2. For each class diagram attribute that is passed as input to this template the template generates a Java variable declaration with `public` visibility. For instance, by passing the AST of the class diagram attribute "int attributeName = 5;" the template generates the Java variable declaration statement "public int attributeName = 5;". Here, the value 5 is the default value. Line 1 of List. 2 generates the variable declaration and line 2 generates the variable instantiation by checking if a value has been defined in the input model. Finally, a semicolon is used to close the Java variable declaration.

```FreeMarker
public ${ast.printType()} ${ast.name}
  <#if ast.value??> = ${ast.printValue()} </#if>
;
```

List. 2: The (simplified) template for generating a Java attribute.

### 4.1 Unit Testing Templates

To present the testing concepts for template-based code generators, we extended the JUnit testing framework to support the different testing approaches for code generator templates. Subsequently, we introduce the resulting TUnit and the realized concepts for early stage unit testing templates.

```Java
@RunWith(de.se.rwth.tunit.TUnitRunner.class)
@TemplateUnderTest(templateName="JavaAttribute",
  type = ASTCDAttribute.class)
public class TUnitTestClass {
  @Test
  @InputModel(fileName = "src/test/" +
    "resources/input/JavaAttribute.cd")
  public void templateTestMethod() {  ...  }
}
```

List. 3: Skeleton of a TUnit test class with one test method.

In JUnit, test runners are used to execute the test methods implemented in a test class. As TUnit introduces custom annotations that are used for configuration purposes, the default JUnit runner is not appropriate to execute the template tests properly. Due to this, TUnit integrates its own test runner that is aware of the semantics of the annotations and knows how to execute the template tests. Thus, each TUnit test class has to be annotated with the TUnit specific test runner as shown in line 1 in List. 3. This listing shows a complete skeleton for a simple TUnit test class.

Three crucial aspects that are relevant when testing template-based generators are: *Which template is under test? Which input model is used for the template under test? Which parts of the input model are relevant for the test?* TUnit introduces two annotations that define the template under test and the test input model.

The mandatory annotation `@TemplateUnderTest` is used to define the template under test and is used for all test methods of the test class. The annotation provides two mandatory attributes: `templateName` and `type`. In `templateName`, the path to the template under test and its name have to be defined. In `type`, the type of the AST node that is handled by the template has to be stated so that the template is only invoked for the specified type. When executing a template test, each test method is executed by first parsing the specified input model. The created AST is then traversed and the template under test is executed for each AST element that is of the specified `type`. Finally, the test method is actually called. List. 3 shows the TUnit test for the `JavaAttribute` template. This template is invoked for AST nodes of type `ASTCDAttribute`.

The input model – defined with the mandatory `@InputModel` annotation – can either be defined on test class level, then the given input model is used in all test methods, or on test method level, then the input model is only used for that particular test method. In each test method, the template under test will be applied for the corresponding input model.

### 4.2  Referencing Generated Output by Model Elements

Defining which template and model elements are under test is the first step to test templates. A further essential step in testing templates is to validate that the template output meets the testers expectations. In JUnit, such expectations are expressed using assert methods. For instance, the assert method `assertEquals` ensures that two values are equal or the assert method `assertNotNull` ensures that a specific value is not null.

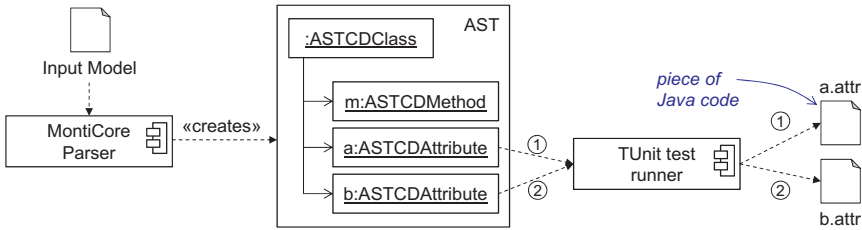

Fig. 5: For each element that fits the specified type of the TUnit test, the template is called and the output is stored in a file.

A prerequisite for being able to formulate such assert statements is that the tester can access the output produced by the template. As explained in Section 4.1, the input model is first parsed when executing a test method. The resulting AST is then traversed and the template is executed for each AST element of the specified type. The output of each template application is stored individually in a distinct file. As the input model for the template may contain multiple elements of the specified type, it is possible that multiple output files are created when executing one test method. Figure 5 shows how TUnit handles the output generated by a template. In this figure, it is assumed that the input model is a class diagram consisting of methods and attributes and that the template under test is defined for AST nodes of type `ASTCDAttribute`. According to Figure 5, the input model contains a class with two attributes. As a result, TUnit creates one file which contains the output of the template application to the first attribute (*a.attr*) and another file for the output of the template application to the second attribute (*b.attr*).

In a concrete test case, a tester usually wants to validate the expectations concerning a specific output, e.g. the output produced for the first attribute in Figure 5. To accomplish this, one option for a tester would be to construct the name of the output file by himself. This is possible as TUnit creates the output files according to a specific naming convention. However, a disadvantage is that it becomes more laborious to define tests. Moreover, the names of the output files will change as soon as the input model will be updated. Testers would need to update the statically referenced output files after each input model update.

To cope with this problem, TUnit (a) traces which output file was created for which AST element and (b) provides an API that allows to retrieve a particular AST element and that returns the corresponding generated file. In Figure 5, the traceability is depicted by the numbers. Thus, the testers can use the API to uniquely identify a specific AST element and the generated file is returned without expecting the testers to construct the concrete path to the output file on their own. Currently, this API is restricted to class diagram input models. Additionally, the API can only be used to address single model elements only, i.e., a model element can be specified in a fully qualified way.

In order to create a test case for a generator template, we subsequently present two assertion variants that both rely on defining the complete expected output.

### 4.3  Assertions for Code Generator Templates

The most basic approach is to perform a simple string comparison between the actual output and an expected string. The tester has to define the complete string that is expected as a result of the template application. A disadvantage of this approach is that the testers are forced to denote the complete expected string, which can be quite laborious and error-prone. Moreover, this approach is rather fragile, as every two varying characters will result in a failing test, e.g., whitespace issues. To cope with the latter problem, TUnit offers a more flexible string comparison method which can be configured to neglect specific types of differences, e.g. differences concerning tabulator characters or indentation.

A more advanced method of creating assertions is to perform an AST comparison. In the course of this comparison, it can be ensured that two AST nodes are (not) equal by including not only the AST node itself but also children elements of the AST node. For this purpose the tester has to define the expected output, which needs to be parsed to build the corresponding AST. Moreover, the template output needs to be parsed to build the AST as well. The AST comparison can then be performed based on these two ASTs. It has to be taken into account that the template output can contain only parts of complete files, e.g. a variable declaration. Due to this, a prerequisite for this approach is a parser for the target language and target language constructs.

Comparing two ASTs means to traverse both ASTs and compare the contained objects. Figure 6 illustrates the comparison of a generated and an expected AST. The one on the left-hand side has been generated by the FreeMarker template shown in List. 2 for the class attribute `int attributeName = 5;`. The AST on the right-hand side of Figure 6 is the AST which was built out of the expected template output. An AST comparison of the generated and the expected AST will reveal the unmatching parts. In Figure 6, this is the variable name and the variable type. As a result, the TUnit assertion will report an error indicating this. A side effect of an AST-based comparison is that the AST of the target language is at hand. This AST can be used to check target language context conditions that may check, e.g. if a variable has been defined before usage. In this way a primary step towards semantically checking the generated code is performed.

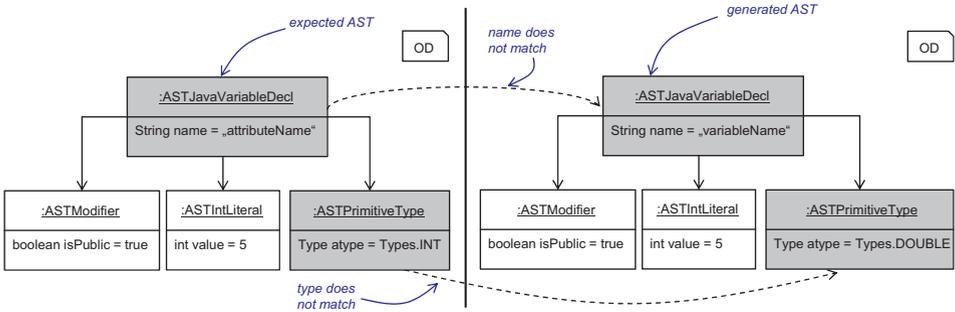

Fig. 6: A comparison of the generated AST (left) and the expected AST (right).

By explicitly stating which template is under test, which input models are used for testing and how the assertions should be handled, template unit testing can be enabled up to the point when the template's context is of relevance for the test. In the following section, the challenge of testing templates that rely on context information is addressed.

## 5   Context-Aware Unit Testing Code Generator Templates

A code generator template that is under test is not always fully self-contained and thus independent of the template engine context. In other words, it requires certain inputs or values to be accessible during execution. For MontiCore code generators such a context may contain variables, helpers, symbol table entries, and template references. Figure 7 shows the same template hierarchy as Figure 4 but the template under test changed to the `JavaMethod` template, which needs extra context information.

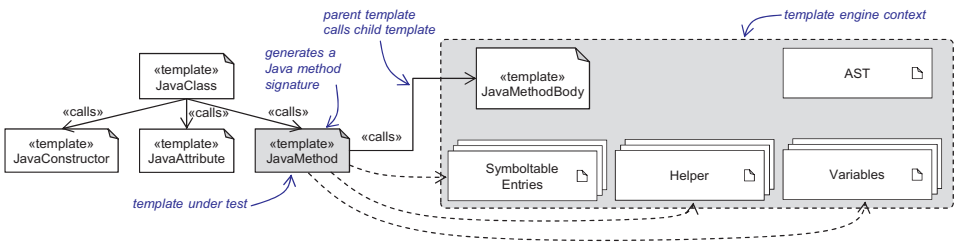

Fig. 7: Testing a code generator template with context.

```
                                                              FreeMarker
1 //Variables: paramType, paramName
2 //Helpers: methodHelper
3 public ${ast.printReturnType()} ${ast.printName()}
4   (${paramType} ${paramName})
5 ${methodHelper.printThrowsDecl(ast)}
6 ${tc.include("cd2data.core.templates.JavaMethodBody", ast)}
```

List. 4: The (simplified) template for generating a Java method.

Assuming that the template under test is the `JavaMethod` template, List. 4 shows the FreeMarker source code, which is an extended version of the template excerpt shown before in List. 1. For example, to generate a Java method the template `JavaMethod` is executed with the input "`void methodName(String param){};`", which is stated in the class diagram. The variable `ast` is used to access the elements of the method declaration - in this case the return type of the method and the method name. The parameter type and parameter name are passed to the template as variables. Additionally, the helper `methodHelper` is used to print Java throws declarations. An instance of this helper is passed as well to the template. In addition, a sub template (see line 6 in List. 4) is called to print the body of the method.

While variables, helper, and symbol table entries of a template under test can easily be mocked to provide enough context for the template to be executed in a test, mocking template references influences the depth of the test with respect to the template hierarchy, i.e., the more templates are mocked, the less templates of the overall template hierarchy are tested. For instance, the `JavaMethod` template, which is currently under test, references the `JavaMethodBody` template, i.e., this sub-template is called and its generated code is embedded in the generated code of the parent template. We refer to the mocking of sub-templates as pruning the sub-templates of the template under test.

### 5.1  Mocking Helpers and Template Variables

In order to mock calls to helper methods, TUnit provides the annotation `@InitHelpers`. This annotation can be used to annotate at most one method in the test class and TUnit expects this method to return a map of strings as keys and objects as values. The strings denote the names of the helper variables and the objects the associated instances of the helper classes. Thus, the tester can define the object to be used when accessing a particular helper variable. He can also implement mocks for helper classes and assign mock objects to the helper variables.

A template can rely on multiple variables that need to be set when calling that template. TUnit supports mocking of variables by providing the annotation `@InitVariables`. At most one method in the test class can be annotated with `@InitVariables` and this method must return a map of strings and strings. The keys of this map denote the variable names to be mocked. The associated values will be used as the variable value when calling the template. In this way, the tester can easily define values for variables needed by a template.

### 5.2 Mocking Symbol Table

As the symbol table stores information about referenced symbols and is part of the code generator template context, it needs to be mocked for testing as well. For mocking symbol tables, TUnit provides the `@SymbolTablePath` annotation for each test class. The overall idea is to provide a set of symbol table models to define all references that are possible and then to create a test model referencing these symbols.

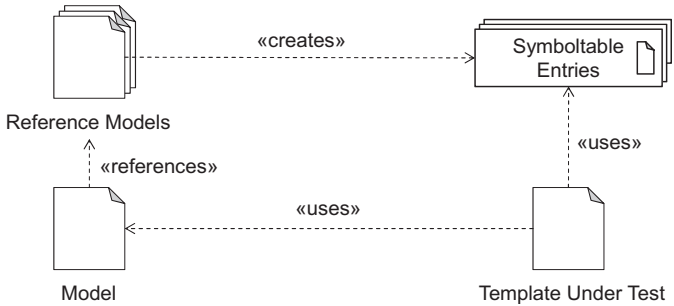

Fig. 8: Overview of the symbol table mocking approach.

In order to extend or mock the symbol table for testing, the testers need to create one or multiple models conforming to the input language of the code generator template, e.g., class diagram language. By annotating a TUnit test class with the `@SymbolTablePath` annotation, the path to the input models that should be used for building the set of symbol table entries is defined. TUnit loads each model and stores all symbols in one symbol table that is provided to the code generator template during execution. An overview of this approach is presented in Figure 8.

This approach of providing symbol table entry information to the template under test is inline with the TUnit's overall approach to separate context information that need to be provided and defining inputs for the template under test. Consequently, context information and in particular symbol table models can be reused for varying inputs.

### 5.3 Mocking Sub-Template Calls

A TUnit test case may not fail for all defined inputs but the overall code generator may still produce invalid code. This is due to embedded sub-template calls in templates under test. The mentioned example of the invalid code produced by the code generator may happen if the embedded sub-template calls are mocked. In contrast, without pruning the sub-templates, creating the TUnit test may be time-consuming, because all helpers, variables, symbol table entries, and template references need to be considered. Clearly, without pruning any sub-templates, the test coverage, i.e., the amount of templates that are executed in one TUnit test, is higher.

When to prune sub-templates depends on the template and the testing strategy. A testing strategy that can be used is to always try to neglect pruning sub-templates if the sub-templates do not generate a crucial part of the overall generated code. If the sub-templates are crucial, they should be pruned and tested in a separate TUnit test. Obviously, the term crucial depends on the tester and the context.

To allow a tester to test templates in isolation, i.e., by abstracting away from the results of sub-templates, TUnit provides the annotation `@TemplateSubstitutionPolicy` that has to be defined at the test class level. With this annotation, the tester can configure the strategy on how to mock sub-template calls:

- *Replace with empty*: Every sub-template call is replaced with the empty string. This imitates the situation that no sub-templates are called at all.
- *Replace all with template*: Instead of calling the sub-templates, every time a self-defined template is called. The output of applying this template is inserted instead of the original template.
- *Replace with string*: In this case, a string is defined that is returned instead of the result of calling the sub-templates.
- *Provide method*: This strategy is the most flexible strategy, as it allows for configuring which specific sub-template call is replaced by which specific string or template. This has to be implemented in a method annotated with `@InitSubtemplates`.

If the template substitution policy is not specified for a test class, the sub-template calls are not mocked and the results of the sub-template calls are inserted into the template output as usual.

## 5.4   Checking Failures with Assertions

One deficiency of the assertion mechanisms presented in Section 4.3 is that the testers have to denote the complete expected output. In case a template generates a large file but only small parts of the output should be checked, applying either of them is too laborious.

In the following, a further variant is proposed, which allows for performing checks for dedicated parts of the AST resulting from the template application. In essence, TUnit provides an API that contains assert methods for different kinds of AST nodes. In the following, a few class diagram specific examples are given:

- `assertHasClass(ASTCompilationUnit, String)`: Ensures that the given compilation unit contains a class with a specific name.
- `assertHasAttribute(ASTClass, String, Type)`: Validates that the given class contains an attribute with the given name and a given type.
- `assertHasMethod(ASTClass, String, Type, List<Type>)`: Ensures that the passed class contains a method with the specified name, return type, and the given list of parameter types.

```Java
//Retrieval of template output omitted here
String testOutputPath = ...

ASTMethodDeclaration actMethodDecl =
  PartialParsing.parseMethodDeclaration(
    new File(testOutputPath));

ASTJavaAssert.assertMethodReturnTypeEquals(
  actMethodDecl, "void");

ASTJavaAssert.assertMethodNameEquals(
  actMethodDecl, "methodName");

ASTJavaAssert.assertMethodHasParameter(
  actMethodDecl, "String", "param");
```

List. 5: Example for AST-based API assertions.

List. 5 shows an example for using the AST-based API. As in the previous example, the actual template output needs to be parsed to create the AST. In the course of this, the parser reports an error, if the code does not represent a valid method declaration. Subsequently, it is at first checked, whether the return type of the parsed method declaration equals the expected return type (line 8 to 9). After that, it is checked that the method name of the parsed method declaration equals the expected name (line 11 to 12) and that the method has a particular parameter (line 14 to 15).

The main advantage of this strategy is that it is usually less laborious to apply it compared to the previously introduced assertion mechanisms as the testers do not have to denote the complete expected result string. Furthermore, this strategy is usually less fragile as the test results are not necessarily affected by every single character change. One potential downside is that the offered API focuses on high-level checks. Hence, it is not well suited to check for all kinds of fine-grained details. Moreover, the API is bound to a particular target language. Consequently, a new API has to be provided in case a new target language is used.

In the presented example of the template under test in List. 4, we have not considered the case that a sub-template may generate a file rather than a string that is embedded in the parent template. These generated files can also be checked with TUnit; however, the testers need to manually consider such "side effects" by manually extending the test to consider the generated artifacts.

## 6 Conclusion and Future Work

The use of code generators in MDD demands for strong testing concepts to develop robust code generators. Most existing approaches to test code generators rely on testing the code generator as a whole, by executing the complete code generator. Testing only selected templates or validating fragments of code is not easily possible in these approaches.

In this paper, we have presented a method and TUnit– an extension of JUnit – as corresponding tool support for testing code generators. It can be employed early in the development of code generators where no complete source code artifacts are generated. Since templates are executed in a context that includes helpers, variables, symbol table references, and template references, TUnit provides means to mock specific parts of this context. TUnit takes input models for a code generator and executes the template under test on selected parts of these models. To validate the template output, three assertion strategies have been presented. First, a string comparison between the actual output and the expected output, which needs to be defined explicitly. Second, an AST comparison based on the input model AST and an expected AST. Third, an AST-based API comparison that allows for executing checks on dedicated parts of the AST that is created from the template output.

Currently, the input model has to be a complete class diagram. In future work we plan to support pieces of class diagrams, e.g. a class only or a method declaration. Besides comparing ASTs to find assertion violations, it is also possible to employ transformation languages. Assuming that a transformation language for the generated language exists [WR11, We12], assertions can be defined by defining patterns that need to be matched in the generated code. If a pattern cannot be found, then the assertion is violated. Otherwise, the assertion is correct. Finally, a general question to be addressed is the efficiency of the proposed approach.

## References

[BKS04]  Baldan, Paolo; König, Barbara; Stürmer, Ingo: Generating Test Cases for Code Generators by Unfolding Graph Transformation Systems. In: Graph Transformations, volume 3256 of LNCS. Springer Berlin Heidelberg, 2004.

[FR07]   France, Robert; Rumpe, Bernhard: Model-Driven Development of Complex Software: A Research Roadmap. In: Future of Software Engineering 2007 at ICSE. IEEE Computer Society, 2007.

[Fr15]   FreeMarker Template Language. http://www.freemarker.org/, October 2015.

[Gr08]   Grönniger, Hans; Krahn, Holger; Rumpe, Bernhard; Schindler, Martin; Völkel, Steven: MontiCore: A Framework for the Development of Textual Domain Specific Languages. In: Companion of the 30th International Conference on Software Engineering. ICSE Companion '08. ACM, 2008.

[Hu11]   Hutchinson, John; Whittle, Jon; Rouncefield, Mark; Kristoffersen, Steinar: Empirical Assessment of MDE in Industry. In: Proceedings of the 33rd International Conference on Software Engineering. ICSE '11. ACM, 2011.

[Jö13]    Jörges, Sven: Construction and Evolution of Code Generators - A Model-Driven and Service-Oriented Approach, volume 7747 of LNCS. Springer, 2013.

[JU15]    JUnit Website. http://junit.org/, October 2015.

[KRV08]   Krahn, Holger; Rumpe, Bernhard; Völkel, Steven: MontiCore: Modular Development of Textual Domain Specific Languages. In: Objects, Components, Models and Patterns. volume 11 of Lecture Notes in Business Information Processing. Springer Berlin Heidelberg, 2008.

[KRV10]   Krahn, Holger; Rumpe, Bernhard; Völkel, Steven: MontiCore: A Framework for Compositional Development of Domain Specific Languages. International Journal on Software Tools for Technology Transfer, 12, 2010.

[Li14]    Liebel, Grischa; Marko, Nadja; Tichy, Matthias; Leitner, Andrea; Hansson, Jörgen: Assessing the State-of-Practice of Model-Based Engineering in the Embedded Systems Domain. In: Model-Driven Engineering Languages and Systems, volume 8767 of LNCS. Springer International Publishing, 2014.

[Ra10]    Rajeev, A. C.; Sampath, Prahladavaradan; Shashidhar, K. C.; Ramesh, S.: CoGenTe: A Tool for Code Generator Testing. In: Proceedings of the IEEE/ACM International Conference on Automated Software Engineering. ASE '10. ACM, 2010.

[Sa08]    Sampath, Prahladavaradan; Rajeev, A. C.; Ramesh, S.; Shashidhar, K. C.: Behaviour Directed Testing of Auto-code Generators. In: Proceedings of the 6th IEEE International Conference on Software Engineering and Formal Methods. SEFM '08. IEEE Computer Society, 2008.

[Sc12]    Schindler, Martin: Eine Werkzeuginfrastruktur zur Agilen Entwicklung mit der UML/P. Aachener Informatik Berichte, Software Engineering. Shaker Verlag, 2012.

[St06]    Stürmer, Ingo: Systematic Testing of Code Generation Tools - A Test Suite-oriented Approach for Safeguarding Model-based Code Generation. PhD thesis, Department of Compiler Construction and Programming Languages of the Technical University of Berlin, 2006.

[St07]    Stürmer, Ingo; Conrad, Mirko; Doerr, Heiko; Pepper, Peter: Systematic Testing of Model-Based Code Generators. IEEE Transactions on Software Engineering, 33(9), 2007.

[SWC05]   Stürmer, Ingo; Weinberg, Daniela; Conrad, Mirko: Overview of Existing Safeguarding Techniques for Automatically Generated Code. In: Proceedings of the 2nd International Workshop on Software Engineering for Automotive Systems. SEAS '05. ACM, 2005.

[We12]    Weisemöller, Ingo: Generierung domänenspezifischer Transformationssprachen. Aachener Informatik Berichte, Software Engineering. Shaker Verlag, 2012.

[WR11]    Weisemöller, Ingo; Rumpe, Bernhard: A Domain Specific Transformation Language. In: ME 2011 - Models and Evolution. 2011.

[Ze06]    Zelenov, Sergey V.; Silakov, Denis V.; Petrenko, Alexander K.; Conrad, Mirko; Fey, Ines: Automatic Test Generation for Model-Based Code Generators. In: Proceedings of the 2nd International Symposium on Leveraging Applications of Formal Methods, Verification and Validation. ISOLA '06. IEEE Computer Society, 2006.